\documentstyle[12pt]{article}
\catcode`\@=11

\@addtoreset{equation}{section}
\def\theequation{\arabic{section}.\arabic{equation}}

\catcode`\@=11

\def\section{\@startsection{section}{1}{\z@}{3.5ex plus 1ex minus
   .2ex}{2.3ex plus .2ex}{\large\bf}}

\def\eqnarray{\let\@currentlabel=\theequation\refstepcounter{equation}
    \global\@eqnswtrue
    \global\@eqcnt\z@\tabskip\@centering\let\\=\@eqncr
  $$\halign to \displaywidth\bgroup\@eqnsel\hskip\@centering
   $\displaystyle\tabskip\z@{##}$&\global\@eqcnt\@ne
      \hfil${{}##{}}$\hfil
      &\global\@eqcnt\tw@ $\displaystyle\tabskip\z@{##}$\hfil
       \tabskip\@centering&\llap{##}\tabskip\z@\cr}

 \def\lefteqn#1{\hbox to 4\arraycolsep{$\displaystyle #1$\hss}}

\newcommand{\req}[1]{(\ref{#1})}
\newcommand{\be}{\begin{equation}}
\newcommand{\ee}{\end{equation}}
\newcommand{\bea}{\begin{eqnarray}}
\newcommand{\eea}{\end{eqnarray}}

\newcommand{\therefore}{\Rightarrow}

\def\IR{{\hbox{{\rm I}\kern-.2em\hbox{\rm R}}}}
\def\IH{{\hbox{{\rm I}\kern-.2em\hbox{\rm H}}}}
\def\IC{{\ \hbox{{\rm I}\kern-.6em\hbox{\bf C}}}}
\def\IZ{{\hbox{{\rm Z}\kern-.4em\hbox{\rm Z}}}}
\def\ID{{\hbox{{\sl 1}\kern-.3em\hbox{\sl I}}}}
\def\IP{{\hbox{{\sl I}\kern-.2em\hbox{\sl P}}}}

\begin{document}

\begin{titlepage}

\vspace{.5in}
\begin{flushright}
UCD-96-29\\
November 1996\\
quant-ph/9611026 \\
\end{flushright}
\vspace{.5in}

\begin{center}
{\large\bf Coherent State Approach to \\
Time Reparameterization Invariant Systems }
\\

\vspace{.4in}
M.\  C.\  A{\sc shworth} \footnote{\it email: mikea@landau.ucdavis.edu}\\
{\small\it Department of Physics}\\
{\small\it University of California}\\
{\small\it Davis, California 95616 USA}\\
\end{center}

\vspace{.5in}
\begin{center}
{\bf Abstract}\\
\end{center}

\begin{center}

\begin{minipage}{5in}
{\small
For many years coherent states have been a useful tool for understanding
fundamental questions in quantum mechanics.  Recently, there has been
work on developing a consistent way of including constraints into the
phase space path integral that naturally arises in coherent state
quantization. This new approach has many advantages over other
approaches, including the lack of any Gribov
problems, the independence of gauge fixing, and the ability to
handle second-class constraints without any ambiguous determinants.
In this paper, I use this new approach to study some examples
of time reparameterization invariant systems, which are of
special interest in the field of quantum gravity.
}

\end{minipage}
\end{center}
\end{titlepage}
\addtocounter{footnote}{-1}

\section{Introduction}

The coherent state formulation of the path integral \cite{KlauderS}
has many advantages over a conventional Feynman path integral.
Because coherent states describe a minimum uncertainty wave packet,
there is a natural relation between the classical system and
underlying quantum system. The coherent state path integral is
intrinsically superior with regards to a canonical coordinate
transform which just amounts to  relabeling of
the states \cite{KlauderG}.  Moreover, for
coherent state path integral, it is possible to find a well regularized
path integral measure (the pinned
Weiner measure) \cite{Klauder5}.

Recent work \cite{Ashworth} - \cite{KlauderC} has included constraints
into this formulation. The first approach was to consider a
semi-classical constraint \cite{Ashworth}.
This constraint can be inserted
by hand into each time step in the construction of the path integral.
The result is the formal path integral where the action is dependent
on the total Hamiltonian.  Klauder \cite{NKlauder1} constructed
a projection operator that maps states defined on the full phase space
onto physical states.
The resulting path integral is independent
of the functional form of the Lagrange multiplier term and hence gauge
invariant.  Later, Klauder and Shabanov \cite{KlauderC}
generalized this approach to a coordinate-free formulation.

For constrained system, coherent states offer
further advantages.  Because the path integral is regularized, we
are not required to gauge fix to remove the infinite volume term
that normally appears. The result is an averaging over the gauge
orbits.  Without gauge fixing, there are no potential problems with
Gribov ambiguities.
Also we are not required to eliminate second-class
constraints \cite{NKlauder1}, nor is there
the possibility of an ambiguous determinant
for this case.

In this paper, we will review coherent state
quantization of constrained system and compare the results of the
author's  semi-classical construction of the
path integral \cite{Ashworth}
with Klauder's projection operator approach \cite{NKlauder1}.
Then we will work through details of two time-reparameterization
invariant systems.  The first example will be the
single harmonic oscillator.  Although this is a very simple example
it can give us some insight into the details of this formulation.
The second example is the double harmonic oscillator.

The double harmonic oscillator is an important
example in the study of quantum gravity \cite{Lawrie}.
This is a good toy model to help understand the
``problem of time.''\footnote{In a time-reparameterization
invariant system, such as
quantum gravity, the roll of the local time
coordinate is difficult to understand. For more about
this ``problem of time,'' Rovelli has written a series of papers
\cite{Rovelli}. }
In essence, one harmonic oscillator can be
used as a quantum ``clock''
to measure the other oscillator.
In terms of coherent state
quantization, the double harmonic oscillator
show the importance of the geometry on the
system \cite{KlauderR} - \cite{KlauderR2}.  The geometry of the
phase space determines the natural kinematical operator
in which the system should be quantized.  For this case, the resulting
reduced phase space is spherical so the kinematical operators
are spin-like operators.  Also this system has
a potential Gribov problem that results in a difference between the
ground state energies of the reduced and Dirac quantizations.
This Gribov problem results because the gauge orbits form a twisted
bundle over the constraint surface.

\section{Coherent state quantization}

The ordinary phase space has a natural Heisenberg-Weyl algebra structure
that comes from the symplectic structure.  This
operator algebra can be used to
construct a coherent state representation of the Hilbert space that
is labeled by the classical phase space coordinates. So,
we will begin by considering a set of $M$ pairs of Heisenberg
operator $\{ \hat {P}_j,\hat{Q} ^k \}$.  These operators obey the standard
Heisenberg-Weyl commutation relations,

\be
[ \hat {P}_j, \hat{Q}^k ] = - i \hbar \delta _{j}^{k}
\qquad
j,k= 1, \ldots ,M .
\ee

\noindent
The coherent state representation is then a
unitary representation of the Heisenberg-Weyl group
acting on some fiducial vector $| \eta \rangle $
chosen from the Hilbert space.

\be
|p,q \rangle = e^{-if(p,q)} e^{-{i \over \hbar} p \cdot \hat Q}
e^ {{i \over \hbar} q \cdot \hat P} | \eta \rangle .
\label{def1}
\ee

\noindent
In most cases, the fiducial vector is chosen such that
the coherent state is ``physical centered,''
$\langle \eta | \hat{P} _j | \eta  \rangle =
\langle \eta | \hat{Q} ^k | \eta \rangle = 0 $.
For this reason, we will choice the fiducial vector to be
the ground state of the harmonic oscillator,
$| \eta \rangle = |0 \rangle $.
This set of states
does not form an orthonormal basis, as is seen in the overlap
function,

\be
\langle p',q' | p, q \rangle =
\exp  \left\{ { -{ 1 \over 4 \hbar}
\left[  | p'-p | ^2 + |q'-q|^2 \right]
+ {i \over 2 \hbar}
\left[ p' \cdot q - p \cdot q'  \right] } \right\} .
\label{pqover}
\ee

\noindent
However, for any choice of the fiducial vector $| \eta \rangle $,
they do admit a resolution of unity,

\be
\ID = \int |p,q \rangle \langle p,q | \prod_{j = 1}^N
{dp_j dq^j \over 2 \pi} .
\label{qpres}
\ee

\noindent
In addition, these states form an
(over)complete set of states on the Hilbert space.
We can represent any vector in our Hilbert space as
a function of ($p,q$) by defining the function to be
$ \psi(p,q) \equiv \langle p, q | \psi \rangle $.
The overlap function , for example eqn.\req{pqover},
is the reproducing kernel
${\cal K} (p',q';p,q) \equiv
\langle p',q' | p,q \rangle $ on the Hilbert space.
This reproducing kernel has the following properties:

\begin{eqnarray}
\psi (p',q') & = & \int {\cal K} (p',q';p,q) \ \psi(p,q)
\prod_{j = 1}^N {dp_j dq^j \over 2 \pi} ,
\\[1ex]
{\cal K} (p'',q'';p,q) & = &
 \int {\cal K} (p'',q'';p',q') \  {\cal K} (p',q';p,q) \prod_{j = 1}^N
{{dp'}_j {dq'} ^j \over 2 \pi} .
\label{qprep}
\end{eqnarray}

We can construct a path integral based on this representation of the
Hilbert space. Unlike the normal path integral which integrates
over the configurations space $(q)$, the coherent state path
integral naturally
integrates over the phase space $(p,q)$.
To construct this path integral, let us
start with the Hamiltonian evolution between two
states $|p,q \rangle$ and $ |p',q' \rangle$.  The matrix
element may be broken in $N+1$ time steps.  Then at each time
step we can insert a resolution of unity.

\begin{eqnarray}
 \langle p',q' |  e^{ -{i \over \hbar} \hat {H} T}  | q,p \rangle
& = &  \langle p',q' | e ^{- {i \over \hbar} \hat {H} (T - \varepsilon)}
\ \ID \ e^{-{i \over \hbar} \hat {H} (\varepsilon) } | q, p \rangle
\nonumber \\
&& \hskip .45in \overbrace{\int d \mu (p_1,q_1)
| p_1,q_1 \rangle \langle p_1,q_1 | }
\nonumber \\[1ex]
&& \hskip 1.2in \vdots
\nonumber \\[1ex]
& = &  \left(
\int \ldots \int \prod_{n=1}^{N} d \mu (p_n,q_n) \right)
\prod_{n=0}^{N} \langle p_{n+1},q_{n+1} |
e^{- {i \varepsilon \over \hbar} \hat {H}}
| p_n,q_n \rangle ,
\nonumber \\[1ex]
|p',q'\rangle  = |p_{N+1}, q_{N+1} \rangle ,
&&\qquad |p,q\rangle  = |p_{0}, q_{0} \rangle ,
\qquad  \qquad \varepsilon = (t' - t) / (N + 1) .
\label{path1}
\end{eqnarray}

\noindent
The measure $d \mu(p,q)$ is the same as the measure defined in the
resolution of unity \req{qpres}.
In the limit ($\varepsilon \rightarrow 0$),
if the paths are reguarded as continuous and differentiable, then
we can formally rewrite the above matrix element
\req{path1} in form of a path integral (see \cite{KlauderS}
for more details),

\be
\int D \mu (p, q) \exp \left\{ {i \over \hbar } \int \Big(
p \dot q - H(p,q)  \Big) dt \right \},
\label{path3}
\ee

\noindent
where the symbol $H(p,q) = \langle p,q| H | p, q \rangle $.
In the stationary phase approximation, this then leads to the
standard Hamilton's equations of motion.

Unlike ordinary configuration space path integrals, the phase
space path integral can be given a natural regularization
by inserting an additional term into the path integral \cite{KlauderR}.
This is done by changing the measure to a pinned Weiner measure.
This measure originally arose in the study of Brownian motion.
The probability density of a particle undergoing Brownian motion
is governed by the diffusion equation.
The fundamental solution of
the diffusion equation  for a flat metric
is a spreading Gaussian,

\begin{equation}
\rho (t''; t') = {1 \over { 2 \pi \nu (t''- t') }}
\exp \left[ -{(p'' - p')^2 + (q''- q')^2  \over 2 \nu (t''-t')}
\right]  .
\label{fund1}
\end{equation}

\noindent
This solution possesses a semi-group structure with the following
product rule:

\begin{equation}
\rho (t'''; t') = \int d p'' \ d q'' \ \rho (t''';t'')
\rho (t''; t') .
\label{prodt1}
\end{equation}

\noindent
To construct the Weiner measure, we proceed in a similar fashion as
we did with the construction of the path integral.  We can use the
product rule \req{prodt1} repeatedly to break the time in $N+1$ steps.

\begin{eqnarray}
&& \rho (t''; t') = \int \left( \prod _{i=1} ^N
d p_i \ d q_i \right)  \left( {1 \over { 2 \pi \nu  \varepsilon }}
\right) ^{N}
\left( \exp { \sum _{i=0} ^N -{(p_{i+1} - p_{i})^2 + (q_{i+1} - q_i)^2
\over 2 \nu \varepsilon}} \right) ,
\nonumber \\[1ex]
&& (q'',p'') = (q_{N+1}, p_{N+1}),
\hskip .2in
 (q', p')  = (q_0,p_0),
\hskip .2in \varepsilon = (t'' - t') / ( N +1 ) .
\end{eqnarray}

\noindent
Then, in the continuum limit, we have a formal expression for the
Weiner measure,

\begin{equation}
d \mu _W ^\nu (p,q) = {\cal N} e ^{ -{1 \over 2 \nu} \int \dot p ^2
+ \dot q ^2  \ d t }\  D q \ D p
\end{equation}

\noindent
Note that the initial and final points of the paths are fixed
(or pinned) on the phase space.
Writing this in a more general way to include other choices of the
metric and higher dimensions, the measure is

\begin{equation}
d \mu _W ^\nu (p,q) =  {\cal N} e ^{-{1 \over 2 \nu} \int
 \left( { d \sigma (p,q)  \over d t} \right) ^2 d t }
\prod_{j=1}^N {D p^j D q^j} .
\label{Weiner}
\end{equation}

\noindent
The measure in \req{path3} can now be replaced by the formally
well defined pinned
Weiner measure just by the addition of the extra factor,

\begin{equation}
e^{ - {1 \over 2 \nu} \int \left( {d \sigma \over dt} \right)^2 dt}.
\end{equation}

\noindent
In the limit $\nu \rightarrow \infty$, this term formally becomes
unity and we
are left with our original path integral \req{path3}.
Unless the action is explicitly dependent on the
the measure of the phase space,
this Weiner measure is the only place that the
geometry of the phase space come into play in the path integral.
This geometry (as we will see in our second example) determines the
natural kinematical operators in which the
system should be quantized (see \cite{KlauderR}).

\section{Constrained Coherent State}

In this section, I will review the methods for applying
first class constraints to coherent state path integrals in general,
and then restrict to the case of time-reparameterization invariant
systems.

To begin with, let us consider a $2M$ dimensional phase space labeled by
coordinates $ (p_i , q^j) $ where $i,j = 1, \ldots M$.
On this phase space, the constraint surface can be defined in terms
of a system of $N$ equations $\phi_a (p,q) = 0$, where $(2M- N)$ is the
dimension of the constraint surface. The evolution on the constraint
surface is generated by the total Hamiltonian,

\be
H_T (p,q) = H(p,q) + \lambda ^a \phi_a (p,q),  \qquad
{d F \over dt}  = \{ F ,H_T \} \Bigg| _{\phi_a = 0} \ .
\ee

For this paper,
we are only interested in the dynamics of a system
where the constraint functions, $\phi _a = 0$,
 are all first class functions,

\begin{equation}
\{ \phi_a , \phi_b \} \approx 0 , \qquad  a,b = 1, \ldots ,N .
\end{equation}

\noindent
We will assume that the time derivatives will not introduce
any new (secondary) constraints. Therefore the set $\{ \phi_a \}$
is complete.  This also means that the Hamiltonian is also a first
class function,

\begin{equation}
{d \phi_a \over d t} =
\{ \phi_a ,  H_T \} = \{ \phi_a ,  H \} + \lambda^b
\{ \phi_a , \phi_b \} \approx 0 .
\label{tham}
\end{equation}

\noindent
Furthermore, because these commutators are all weakly vanishing,
near the constraint surface,
they are given as linear combinations of the
constraint functions \cite{Henn1}.
Thus the Hamiltonian and the constraints form a closed algebra,

\bea
\{ H , \phi_a \}  & = & h_a^c \phi_c ,
\label{algebra2} \\[1ex]
\{ \phi_a, \phi_b \} & = & C _{ab}^c \phi_c .
\label{algebra1}
\eea

We can now go on to study solutions to the
time evolution equation on our constraint surface,

\begin{equation}
{d f \over dt} = \{ f, H_T \} =
\{ f,H \} + \lambda^a \{ f, \phi_a \} .
\label{evolt}
\end{equation}

\noindent
We see that in general the solution to eqn. \req{evolt}
will depend on the choice of the Lagrange
multiplier $\lambda ^a$.  However, a physical observable will not have
any dependency on this choice.  Therefore, any solution that
differs only by changing the value of
the Lagrange multiplier is defined to be
equivalent.
For first class constraints,
these gauge transformation are generated by
the constraint equations \cite{Henn1},

\begin{equation}
\delta f =
\delta \varepsilon ^ a \{ f , \phi_a \} ,
\end{equation}

\noindent
and the dimension of the gauge transformations
is the same as the number of constraints.
Thus the reduced phase space (the manifold after applying
the constraints and quotienting out the gauge orbits) is then
a ($2M - 2N$) dimensional manifold.  This reduced phase space admits
a local symplectic structure \cite{Henn1}.
So, it is possible to locally find a canonical coordinate
system such that

\be
\{ \tilde p_i, \tilde q^j \} = \delta ^j_i ,
\qquad i,j = 1, \ldots , N.
\ee

Let us consider, these two phase space.  In general we may describe the
dynamics of the system on either the full phase space in terms of the
total Hamiltonian or on the reduced phase space in terms of the
reduced Hamiltonian,

\begin{equation}
H_0 = H(p,q) \Big| _{\phi_a =0} \ .
\end{equation}

\noindent
So we already have the a set of coherent states on the the full
phase space,

\be
|p,q \rangle = e^{-if(p,q)} e^{-{i \over \hbar} p \cdot \hat{Q}}
 e^{{i \over \hbar} q \cdot \hat{P} }
| \eta \rangle,
\label{cstate1}
\ee

\noindent
where $ \hat{P},\hat{Q}$
obeys the standard Heisenberg-Weyl commutation relations
$[\hat{P}_j ,\hat{Q} ^k ] = i \hbar \delta ^k_j$.
So now, let us take the naive approach that we can construct a coherent
state on the reduced phase space in the following way. We
can try to use the symplectic structure on the reduced phase space to
define an other set of Heisenberg-Weyl operators
$[ \hat P _j^{~ \prime} , \hat Q  ^{k ~ \prime} ] = i \hbar \delta ^{jk}$
where $j,k = 1, \ldots ,N$.  Note that these operators may not
be global well defined nor are they necessarily defined in terms
of the Heisenberg operators from the full phase space.
For a least for some covering space of
a large patch, we can construct the coherent state,

\be
|\tilde p, \tilde q \rangle =
e^{-i \tilde f( \tilde p, \tilde q)}
e^{-{i \over \hbar} \tilde p \cdot \hat{ Q '} }
e^{ {i \over \hbar} {\tilde q} \cdot \hat{ P '}}
| \eta \rangle .
\label{cstate2}
\ee

\noindent
If the initial and final states,
$| \tilde p,' , \tilde q' \rangle$ and
$| \tilde p'', \tilde q'' \rangle$, can be lifted
back up onto the full phase space, $|p',q' \rangle$
and $| p'', q'' \rangle$, the the resulting dynamics should be
equivalent (up to possible normalizations),

\be
\langle p',q'| e^{- {i \over \hbar} \hat H_T (t'-t)} | p,q \rangle
\sim \langle {\tilde p}',{\tilde q}'|
 e^{- {i \over \hbar} \hat H_0 (t'-t)} | \tilde p, \tilde q \rangle .
\ee

As we can see, there
are two problems that we need to deal with in comparing these
two descriptions. One is that we need to construct a set of
meaning coherent states on the reduced phase space.  We will
see how this is done when we consider the projection approach
of Klauder's \cite{NKlauder1}.  The other problem is
to understand the dynamics of the total Hamiltonian in terms
of a system of coherent states.

To begin with,
let us consider the evolution generated by the
total Hamiltonian \req{tham}.  We can use the resolution of unity
on the full phase space to construct the path integral \req{path1}.
This gives us the path integral,

\begin{equation}
{\cal N} \int D p D q \exp -{i \over \hbar} \left\{ \int
i \hbar \langle p,q | {d \over d t} | p,q \rangle  +
\langle p,q | \hat {H} + \lambda ^a  \hat \Phi _a | p,q \rangle
\right\} .
\end{equation}

\noindent
We can  replace the operators in terms of
either the ``upper'' or ``lower'' symbol depending on our
construction of the path integral
(see  \cite{KlauderS} for more about these symbols).
The resulting path integral is then

\begin{equation}
{\cal N} \int D p D q \exp -{i \over \hbar} \left\{ \int
p \dot q - H(p,q) - \lambda ^a \phi_a(p,q)
\right\} .
\end{equation}

\noindent
We see that the results of the
path integral depends on the choice of the Lagrange
multipliers $\lambda^a(t)$.  To fix this, we can extend our phase space to
include the Lagrange multipliers (see \cite{Klauder3} for
an example of such an extendition).
or reconstruct the path integral
by placing the constraints in at each time step by hand \cite{Ashworth}.
The resulting new path integral includes integrating over the
Lagrange multiplier,

\begin{equation}
{\cal N} \int D p D q D \lambda
\exp \left\{ -{i \over \hbar}  \int
p \dot q - H(p,q) - \lambda ^a \phi_a (p,q)
\right\} ,
\end{equation}

\noindent
and when we integrate over $\lambda ^a$, the result is our
constraint equation,

\be
\int D \lambda \exp \left \{ \lambda ^a \phi_a \right \}
= \delta ( \phi_a ) .
\ee

Another method of imposing the constraints is to project the
coherent states onto the physical states \cite{NKlauder1}.
This projection operator for first class constraints
commutes with the time evolution.
Thus a physical state will evolve into another physical
state.  When this projection operator is included, the
resulting path integral picks up an additional term which is
just a normal integration over the Lagrange multiplier.
The resulting path integral becomes independent of the
functional form of the Lagrange multiplier.

In either case, we are still left to contend with the gauge
degrees of freedom.  Normally, when we integrate over these degrees
of freedom, we will get the volume of the space of paths
for the gauge orbits.  In an ordinary path integral,
this volume term would be infinite and we would have to include
a gauge fixing term to remove this infinite redundancy.
With a coherent state path integral, we can use the Weiner measure to
regularize the path integral, and because of this, we are not
forced to introduce any gauge fixing into the system.
The result is just an well defined averaging over the gauge degree of
freedom.  Then because we not required to gauge fix the system,
we avoid any possible Gribov problem.\footnote{A Gribov problem
or obstruction occurs when the
the gauge fixing term can not be defined globally or that gauge fixing
function intersects a gauge orbit more then once.}

Let us begin a more detailed construction of this path integral
by considering Klauder's projection operator
approach \cite{NKlauder1}.  We wish to find a projection operator
that takes any state onto a state that is annihilated by the
constraint operator (or physical states),

\be
|p,q \rangle _{phys} = \IP |p,q \rangle .
\ee

\noindent
As a standard projection operator $\IP$ must have the following
properties:

\be
\IP^2 = \IP \quad {\rm and} \quad \IP^\dagger = \IP.
\label{proper}
\ee

\noindent
We can construct an example of such a projection operator in terms of the
constraint functions. As we have seen,
the constraint functions form a Lie algebra \req{algebra1}.  Let us
assume that the we can find the corresponding constraint
operators such that
this algebra is carried over to the commutator algebra.
We can use the group elements generated by these operators
to form a projection operator,

\be
\IP = \int e^ {i \lambda ^a \hat \Phi_a} d \mu (\lambda) .
\ee

\noindent
We will choose the measure to be normalized
$\int d \mu (\lambda) = 1$.  In addition, it must satisfy the above
properties of the projection operator \req{proper}.
For a compact group, such a measure is the
left and right equivalent Haar measure (see \cite{NKlauder1}
for more details).  This projection
operator then projects onto the states that obey the
quantum operator equation $ \hat \Phi_a  | \psi \rangle
= 0 $, which are the physical states.

For a non-compact group, finding a measure that
is normalizable is a bit more difficult.
Klauder \cite{NKlauder1} suggested the following idea.
Let the measure take the form,

\be
\IP  = \int e^ {i \lambda \hat \Phi} \left(
{2 \sin \varepsilon \lambda \over \pi \lambda} d \lambda \right).
\label{uncmeas}
\ee

\noindent
This projects onto states where the constraints operator is within
a small interval,

\be
|| \hat \Phi |p,q \rangle _{phys} || \leq
|| \varepsilon \ | p,q \rangle _{phys} || .
\ee

\noindent
Then in the limit
$\varepsilon \rightarrow 0$, we have a handle on how to
regularize this measure.

In order to use this projection operator in our construction of
a path integral, we note that because the measure is left invariant,
the projection operator is invariant under gauge transformations
that are generated by the constraint,

\bea
e^{i \sigma ^a \hat \Phi_a } \IP  & = &
\int e^ {i ({\sigma \cdot \lambda})^a \hat \Phi_a } d \mu( \lambda)
\nonumber \\[1ex]
& = & \int e^ {i {\lambda_a \hat \Phi_a }}
 d \mu(\sigma ^{-1} \cdot \lambda)
\nonumber \\[1ex]
& = & \IP .
\label{gauge1}
\eea

\noindent
In a similar fashion because the addition of the
Hamiltonian operator into the algebra is also closed
(see eqn. \ref{algebra2}), the projection operator
commutes with the time evolution operator,

\be
\IP e^{-{i t \over \hbar} \hat{H}} =
e^{-{i t \over \hbar} \hat{H}} \IP =
\IP e^{-{i t \over \hbar} \hat{H}} \IP .
\label{comtime}
\ee

For the time evolution of the physical states, because the
projection operator commutes with the evolution operator
and the Lagrange multiplier term can be
absorbed into the projection operator, the matrix element can
written in terms of just the evolution of the physical state on the
full phase space.

\bea
\langle p',q' ; t | p,q \rangle _{phys}
& = & \langle p',q' | e^{ - {i t \over \hbar} \hat{H_T} }
\IP | p , q \rangle
\nonumber \\[1ex]
& = & \langle p',q' |e^{-{i t \over \hbar} \hat{H} + i \sigma^a
\hat \Phi_a }
 \IP | p,q \rangle
\nonumber \\[1ex]
& = & \langle p',q' | e^{ - {i t \over \hbar} \hat{H} }
\IP | p , q \rangle
\label{conp}
\eea

\noindent
Then we can place a resolution of unity between the projection and
time evolution operator. After doing this, the first term becomes the
evolution on the full phase space, for which we have already constructed
the path integral \req{path1}.  Formally, we have the following
modified path integral (see \cite{NKlauder1} for more details):

\be
\int \exp \Bigg \{ {{i \over \hbar} \int p \dot q - H(p,q) } \Bigg \}
\langle p'',q'' | \IP |  p,q \rangle
Dp Dq .
\label{path5}
\ee

\noindent
By not absorbing the Lagrange multiplier term into the projection
operator, we can repeat the same process with the Hamiltonian
replaced by the total Hamiltonian operator.  The resulting path
integral is

\be
\int \exp \Bigg \{ {{i \over \hbar} \int p \dot q - H(p,q) -
\lambda^a \phi_a (p,q) } \Bigg \}
\langle p '',q ''| \IP |  p,q \rangle
Dp Dq  .
\ee

\noindent
So even though this path integral appears to still depend on the
choice of the functional form Lagrange multiplier, we see that
in fact it is equivalent to the path integral without this term
\req{path5}.

Now let us consider a different construction of the path
integral.  Let us work with only one constraint $\phi(p,q) = 0$.
We have the projection operator given by

\be
\IP = \int e^{- i \tau \hat \Phi } d \mu ' (\tau).
\ee

\noindent
Let $\tau(t_1,t_2) = \int_{t_1}^{t_2} \lambda (t) dt$. The operators
are inherently time independent. So, we have

\be
\IP = \int e^{- i \int_{t_1}^{t_2}
 \lambda (t) \hat \Phi dt } d \mu ' \Big( \tau(t_1,t_2) \Big).
\label{proj2}
\ee

\noindent
Then using the properties of the projection operator \req{proper},
we can construct the  simple product rule,

\bea
\IP & = & \IP^2
\nonumber \\[1ex]
 & = & \int \int e^{- i \int_{t_1}^{t_2} \lambda
(t) \hat \Phi dt} e^{- i \int_{t_2}^{t_3} \lambda
(t) \hat \Phi dt} d \mu' \Big( \tau(t_1,t_2) \Big)
 d \mu' \Big( \tau(t_2,t_3) \Big)
\nonumber \\[1ex]
 & = &  \int e^{- i \int_{t_1}^{t_3} \lambda (t) \hat \Phi dt }
 d \mu' \Big( \tau(t_1,t_2) + \tau(t_2,t_3) \Big)
\nonumber \\[1ex]
& = & \int e^{- i \int_{t_1}^{t_3} \lambda (t) \hat \Phi dt }
d \mu' \Big( \tau(t_1,t_3) \Big)  .
\label{prule}
\eea

\noindent
This projection operator can be broken into $N$ time segments
as we did in the construction of the path integral \req{path1}.
So we can repeat this construction to include the projection
operator.  The above constrained propagator \req{conp}
then can be written in terms of the discrete path integral.
Let each time step be given by $\varepsilon$, then
$\int_{t_n}^{t_n+1} \lambda (t) dt \approx \varepsilon \lambda_n$
and the measure is $d \mu'(\tau) = d \mu'
(\varepsilon \lambda)$.
The discrete path integral is

\be
\int \ldots \int \prod_{n=1}^{N} d \mu (p_n,q_n)
\prod_{n=1}^{N}  d \mu' (\varepsilon \lambda_n)
\prod_{n=0}^{N} \langle p_{n+1},q_{n+1} |
e^{- {i \varepsilon \over \hbar} \hat{H}}
e^{i \varepsilon  \lambda_n \hat{\Phi}}
| p_n,q_n \rangle .
\ee

\noindent
Let us rescale the Lagrange multiplier
$\lambda \rightarrow \lambda / \hbar$.  Then we see that have derived
the time evolution operator in terms of the total Hamiltonian,

\be
\int \ldots \int \prod_{n=1}^{N} d \mu (p_n,q_n)
\prod_{n=1}^{N} d \mu ' \left( {\varepsilon \over \hbar}
 \lambda_n \right)
\prod_{n=0}^{N} \langle p_{n+1},q_{n+1} |
e^{- {i \varepsilon \over \hbar} \left(
\hat{H} + \lambda_n \hat{\Phi} \right)}
| p_n,q_n \rangle .
\ee

\noindent
In the Continuum limit, we wish to replace the ordinary
measure above with the well defined
Weiner measure.  Certainly, we already know to do this for the first
measure of the momentum and position, but we would also like to do
the same for the Lagrange multiplier measure.

Looking carefully at \req{prule}, we see that in fact we already
have a path integral.  We can also see that if we let the measure
be defined in terms of the fundamental solution of the diffusion
equation \req{fund1}, then we can write the measure as

\bea
&& d \mu' \Big( \tau (t_1,t_2) \Big) = \rho (t_1,t_2)\  d \lambda (t_2)
\nonumber  \\[1ex]
&& \rho (t_1,t_2) = \sqrt{ 1 \over 2 \pi \nu ' (t_1 - t_2) }
\exp \left \{  - { \big( \lambda(t_1) - \lambda(t_2) \big)^2
\over 2 \nu (t_1 - t_2) } \right \} .
\eea

\noindent
We see that this measure is normalized,

\be
\int d \mu (\tau) = \int \rho(t_1,t_2)  d \lambda(t_2) = 1 ,
\ee

\noindent
and the product rule of the this measure \req{prodt1} is consistent
with the above product rule \req{prule}.  Then in the formal limit,
we should replace the measure with a Weiner measure.  Note however that
this Weiner measure is not pinned at both ends, but in fact we should
integrate over the end terms.  This integration is how the
propagator loses its dependence on the Lagrange multiplier.
So this measure can be taken into an unpinned Weiner measure.
Formally we can  write the path integral as

\be
{\cal N}
\int d \mu^{\nu}_W(p,q) d \mu^{\nu'}_W(\lambda)
\exp \Bigg \{ {{i \over \hbar} \int \left( p \dot q - H(p,q)
- \lambda(t) \phi(p,q) dt \right) } \Bigg \} .
\label{path6}
\ee

\noindent
This path integral is discussed in earlier work by the author
\cite{Ashworth} in terms of the semi-classical construction of the path
integral.

Now that we have constructed the various forms of the
path integral for the constrained system (eqs. \ref{path5},
\ref{path6}), we would like to consider
time-reparameterization invariant systems.  A large class of
time-reparameterization invariant systems may be written
in terms of a single constraint, $H_T = \lambda ( \hat{H} - E)$.
A rescaling of the time coordinate can be aborted into the definition
of the Lagrange multiplier.  So $\lambda(t)$ is just a lapse function.

In terms of the coherent state quantization, the matrix element is
quite simple. Because the Hamiltonian is
zero there is no ``time'' evolution on the full phase space,
the matrix element \req{conp} is then just

\bea
\langle p', q' ; t| p ,q \rangle_{phys} & = & \langle p' q' | \IP
| p , q \rangle
\nonumber  \\[1ex]
& = & \int d \mu(\tau)
\langle p '', q'' | e^{ - {i \tau}  {(\hat{H} - E ) \over \hbar} }
| p', q' \rangle .
\label{matrix2}
\eea

\noindent
Because there is no dependence on the position and momentum through
the Hamiltonian, the path integral \req{path6} becomes trival to
integrate in this direction.  The remaining path integral is just
dependent on the Lagrange multiplier.  Then we can use the product
rule \req{prule} to integrate along this direction. The resulting
matrix element is the same as above \req{matrix2}.
Note that the integration variable for this operator should be
identified with
the proper time $\tau = \int_{t_1}^{t_2} \lambda(t) dt$.
This was first noted by Govaerts \cite{Govaerts} in his consideration
of the free particle case.

Let us now consider two examples
of time-reparameterization invariant systems; the single and
double harmonic oscillator.

\section{The single harmonic oscillator}

The harmonic oscillator is a natural place to begin the study of time
reparameterization invariant systems. In addition to being a simple
system to work with it is also the natural setting in which coherent
states first appeared.\footnote{In 1926, Schr\"odinger was interested
in finding a wavefunction for the harmonic oscillator where the center
oscillated at the classical frequency \cite{Schrod}.}
In this section, we will compare the recent
projection operator approach to standard Dirac and reduced phase
space quantization.

We will begin with a quick review of the classical time
reparameterization invariant harmonic oscillator.  The total Hamiltonian
for this model is given by

\be
H_T = \lambda \Big[ {1\over 2} ( p^2 + \omega^2 q^2) - E \Big] .
\ee

\noindent
The action for this system is then

\be
L= \int p dq - \int \lambda \Big[ {1\over 2}
( p^2 + \omega^2 q^2) - E \Big] dt.
\ee

\noindent
The equations of motion can easily be solved by defining the
proper time $\tau = \int_0^t \lambda dt$.  Then the equations of
motion appear as the normal equations of motion for the harmonic
oscillator with $\tau$ replacing the time variable

\be
{d p \over d \tau} = - \omega ^2 q \qquad
{d q \over d \tau} =  p  .
\label{eqmotion1}
\ee

\noindent
The solutions of these equations of motion are

\be
q = A \cos( \omega \tau + \phi) \qquad
p = A \omega \sin (\omega \tau + \phi)  .
\label{eqmotion2}
\ee

\noindent
In addition the equations of motion we also must satisfy the
constraint equation,

\be
{1 \over 2} (p^2 + \omega ^2 q^2) - E = 0  .
\label{singcon}
\ee

\noindent
Substituting the equations of motion \req{eqmotion2}
into the constraint equation, we can solve for the
amplitude,

\be
A = {\sqrt {E} \over \omega} .
\label{amp1}
\ee

The remaining degree of freedom of this system
$\phi$ is just the gauge degree of
freedom. To see this, let
$\lambda \rightarrow \lambda + \varepsilon$, then
$\phi \rightarrow \phi' = \phi + \varepsilon t
+ {\cal O} (\varepsilon^2)$.  So the resulting
reduced phase space is
just a single point.  Quantizing this system is trivial, because
there is only one state. Note, however, that the energy
$E$ appears to be arbitrary.

Now, let us look at the Dirac quantization of this system.
To begin with, we must find the
operator corresponding to the constraint
function.  It is natural to choose a Hermitian operator.
For convenience, we will switch to the complex coordinate
$\alpha = \sqrt{ \omega \over 2 \hbar} q + i
\sqrt{1 \over 2 \omega \hbar} p$, and we will replace the momentum and
position operators by the standard harmonic oscillator raising and
lower operator ($a, a^\dagger$).
The  constraint operator can be written

\be
\hat {\Phi} = {\omega \hbar \over 2} (a a^{\dagger} + a^\dagger a )
 -  E \ID .
\ee

\noindent
Let us define $E' = E / \omega \hbar - 1 / 2$ and rescale $\lambda$.
Then the constraint operator can be written in terms of the number
operator ($a^\dagger a$),

\be
\hat {\Phi} = a ^{\dagger} a - E' \ID .
\label{conop1}
\ee

\noindent
Following Dirac quantization,
the physical states are defined as the states that are
annihilated by the constraint operator.
Therefore, the physical state is an eigenstate of
the number operator.  This also imposes the condition that $E'$
is an integer.

\be
 a ^{\dagger} a | \Psi \rangle_{phys}  =
 E' | \Psi \rangle_{phys}
\qquad \therefore
\qquad
 | \Psi \rangle_{phys} = | n \rangle ,
\qquad E' = n .
\label{dirac1}
\ee

\noindent
The Dirac quantization also leads to the single state $ | n \rangle$.
However it imposes the restriction that the energy is quantized
$E = \hbar \omega (n + 1/2)$.

We would now like to consider this system in terms of coherent states.
We can project the coherent state on the full phase space
$| \alpha \rangle $ onto the
physical space by using the projection operator $\IP$.

\bea
| \alpha \rangle _{phys} & = & \IP | \alpha \rangle
\nonumber \\[1ex]
& = & \int e^{ i \lambda ( a^\dagger a - E')} \delta \lambda
\left( e^ {-{|\alpha|^2 / 2}} \sum _{n=0} ^{\infty}
{\alpha^n \over \sqrt{n!}} | n \rangle \right)
\nonumber \\[1ex]
& = &  e^ {-{|\alpha|^2 / 2}} \sum _{n=0} ^{\infty}
{\alpha^n \over \sqrt{n!}}
\left( \int e^{ i \lambda ( n - E')} \delta \lambda
\right) | n \rangle
\label{calc}
\eea

\noindent
Using the measure for non-compact groups \req{uncmeas},
the integral becomes

\be
\int e^{ i \lambda ( n - E')} \left( {2 \sin \varepsilon
\lambda \over \pi \lambda} \right) d \lambda \quad = \quad
\left\{
\matrix{
1  &\hskip.2in & |E' - n| \ < \ \varepsilon
\cr
1/2   && |E' - n|\  = \ \varepsilon
\nonumber \cr
0              && |E' - n| \ > \ \varepsilon
\nonumber } \right. .
\label{dfunc}
\ee

\noindent
We can choose $\varepsilon$ to be arbitrarily small.
Therefore, we see that
$E'$ must be arbitrarily close to an integer $m$
otherwise the physical vector is null. So, if we let $E'=m$, we can
calculate the above sum \req{calc}.

\be
| \alpha \rangle _{phys} = e^ {-{|\alpha|^2} / 2}
{\alpha^m \over \sqrt{m!}} | m \rangle
\label{uncp}
\ee

\noindent
However, this physical state is not yet normalized in the new space.
After normalizing, the physical state is the energy eigenstate with a
phase factor out in front.

\be
| \alpha \rangle ' _{phys} =
{ | \alpha \rangle  _{phys} \over
| \langle \alpha | \alpha \rangle | _{phys} }
= {\alpha ^m \over | \alpha |^m }
| m \rangle = e^ {i m \theta} |m \rangle .
\label{single}
\ee

\noindent
This phase factor is obviously irrelevant to the physics of this
system, and it is easy to see that, in fact, it is the gauge degree
of freedom generated by the constraint \req{singcon}.  It is clear
that the projection method in this system is equivalent to
the Dirac quantization \req{dirac1}.

The ``evolution'' of this system in the reduced
phase space is trivial since there is only one state.  However
we would like consider the matrix element on the full phase
space so we can compare with the classical solutions.
On the full phase space,
we can look at the physical state on the full space space.
Let the physical state be labeled by

\be
|\phi _{\alpha'} \rangle
= |\alpha ' \rangle _{phys},
\ee

\noindent
where $|\alpha ' \rangle $ is the state that is normalized in terms
of the full phase space \req{uncp}.
The matrix element, which is also the wavefunction on the
full phase space, is then

\bea
\phi_{\alpha'} (\alpha '') =
\langle \alpha '' | \phi _{\alpha'} \rangle
& = & \langle \alpha '' | e^ {-{|\alpha '|^2} / 2}
{{\alpha'}^m \over \sqrt{m!}} | m \rangle
\nonumber \\[1ex]
& = & e^ {-{|\alpha ''|^2} / 2} e^ {-{|\alpha '|^2} / 2}
{( {\alpha'} {\bar \alpha''} ) ^m \over m!}.
\eea

\noindent
Let $\alpha' = r' e^{i \theta'}$ and $\alpha ''= r'' e^{i \theta''}$,
and let us renormalize the wavefunction such that it is
approximately one at the peak, ${r'}^2 = {r''}^2 = m$.

\be
\Big| \phi_{\alpha'} (\alpha '')  \Big| =
\sqrt{ 2 \pi m} \
e^ {-{|r''|^2} /2} e^ {-{|r '|^2} /2 }
{ {r'}^m {r''} ^m \over m!}.
\label{phystate2}
\ee

\noindent
This normalization can be explained
in term of a ``gauge fix''.  We want the phase factor from the
initial physical state \req{single} to be one $  e^{i m \theta'} = 1$.
Then, normalizing this function \req{phystate2} over phase space, we have

\be
\int \Big| \phi_{\alpha'} (\alpha '')  \Big|^2
\Big|  \delta \left( e^{i m \theta'} - 1 \right) \Big|
\left( {d \alpha ' d \bar \alpha ' \over \pi} \right)
\left( {d \alpha '' d \bar \alpha '' \over \pi} \right)
= 1.
\ee

\noindent
So there seems to be a natural choice for the form of the
gauge fixing term in this system.

Ordinarily, we would have to construct a set of gauge invariant
operators to work on the reduced phase space, but in this system, the
gauge orbits are are understood to be the phase of the state.
Because we know the behavior of the gauge orbits, we can remain on
the full phase space and consider the correlation between
the physical state and the other coherent states on this space.
Some of the important correlation functions of this system are

\bea
 \langle \alpha '' | \hat{H} | \phi_{\alpha '} \rangle & = &
 \left( m +{1 \over 2} \right)
 \langle \alpha ''| \phi_{\alpha '} \rangle
 \\[1ex]
 \langle \alpha '' | \hat {Q} | \phi_{\alpha '} \rangle
& = & \sqrt{ 2 \hbar \over \omega}
\left( {{m \over \alpha ''}  + \alpha ''} \right)
\langle \alpha ''| \phi_{\alpha '}  \rangle
\\[1ex]
\langle \alpha '' | \hat {P} | \phi_{\alpha '}  \rangle
& = & {i} \sqrt{ 2 \hbar \omega}
\left( {{m \over \alpha ''}  - \alpha ''}  \right)
\langle \alpha ''| \phi_{\alpha '} \rangle .
\label{corr1}
\eea

\noindent
Once again these correlations are peeked when
the classical classical constraint functions are meet,
$| \alpha ''| = | \alpha '| = m$. Then, the
classical limit (expanding about the peak) gives

\bea
 \langle \alpha '' | \hat{H} | \phi_{\alpha '} \rangle & = &
 m+{1/2} + {\cal O }(\hbar),
 \\[1ex]
 \langle \alpha '' | \hat {Q} | \phi_{\alpha '} \rangle & = &
 {\sqrt{ E } \over \omega} \cos (\theta '' - \theta ')
 + {\cal O }(\hbar),
 \\[1ex]
 \langle \alpha '' | \hat {P} | \phi_{\alpha '}  \rangle &  = &
 {\sqrt{ E }} \sin (\theta '' - \theta ')
 + {\cal O }(\hbar) .
\label{eqmotion4}
\eea

\noindent
Then, we can identify $\theta '' - \theta' = \omega t + \phi $.  The
resulting corrections then give us back the classical equations of motion
(eqs. \ref{eqmotion2}, \ref{amp1}).

\section{Double harmonic oscillators}

Next, we would like to consider a system of two independent
but identical harmonic oscillators. In addition to being a
non-trivial example of a time- reparameterization
invariant system (it still has two degrees of freedom
remaining after applying the constraints),
the double harmonic oscillator has been of interest in helping to
understand the ``problem of time'' in quantum gravity.
One of the oscillators can be thought of as a quantum clock.  Then
the other oscillator can be written in terms of the ``time'' that
this clock reads (see \cite{Lawrie} for more about this system).

In this system, we also encounter a potential Gribov problem.
The constraint surface is topologically a three sphere ${\bf S}^3$.
The gauge orbits are topologically equivalent to
a circle ${\bf S}^1$. The resulting reduced phase space is the
two sphere ${\bf S}^2$.  However the
three sphere is not a trivial bundle over the two sphere,
${\bf S}^3 \neq {\bf S}^2 \times {\bf S}^1$, but rather a
twisted bundle.
Therefore, we can't find a global gauge fixing
condition \cite{Henn1}.  We can find a local gauge fixing and
extend it to cover all but a single point of the gauge
orbit.  How we treat this point will determine the ground state
energy for the reduced phase space quantization.

Once again, let us start by considering the classical system.
We will choose each of the harmonic oscillators to
have the same frequency $\omega_1 = \omega_2 = \omega$.
Then the Hamiltonian for the double harmonic oscillator is given by

\be
H_T = \lambda \left( {1 \over 2} ({p_1}^2 + {\omega} ^2 {q_1} ^2)
+  {1 \over 2} ({p_2}^2 + {\omega} ^2 {q_2} ^2) - E \right) .
\ee

\noindent
The action of this system is given by

\be
S= \int p_1 dq_1 + p_2 dq_2 - \int H_T \ dt .
\ee

\noindent
As in the single harmonic oscillator case \req{eqmotion1},
the equations of motion are
easily solved in terms of the proper time
$\tau = \int _0 ^t \lambda(t) dt$ .

\bea
& q_1 = A \cos( \omega \tau + \phi)  \qquad &
  p_1 = A \omega \sin (\omega \tau + \phi)
\nonumber \\
& q_2 = B \cos( \omega \tau + \phi')  \qquad &
  p_2 = B \omega \sin (\omega \tau + \phi').
\label{eqmotion3}
\eea

\noindent
The constraint equation,

\be
{1 \over 2} ({p_1}^2 + {\omega} ^2 {q_1} ^2)
+  {1 \over 2} ({p_2}^2 + {\omega} ^2 {q_2} ^2)  =  E ,
\label{constr2}
\ee

\noindent
limits the amplitudes to

\be
(A \omega)^2 + (B \omega) ^2  =  E.
\label{constr3}
\ee

\noindent
If $ \lambda \rightarrow \lambda + \varepsilon$, the we see that
the gauge transformation take both
$\phi \rightarrow \phi + \omega \varepsilon t $ and
$\phi' \rightarrow \phi' + \omega \varepsilon t $.  So the degree
of freedom that is independent of the gauge transformation
is the difference between the initial phase of the two harmonic
oscillator $\Delta \phi = \phi - \phi '$.
The resulting reduced phase space is two dimensional.

We can take for the coordinates on the reduced phase the
momentum and position of the first harmonic oscillator.
This set of coordinates inherits the symplectic structure
from the full phase space $\{ q_1, p_1 \} = 1$.  However,
the metric on this reduced phase space is no longer flat.
Let us rescale the the momentum $p_i \rightarrow
\sqrt{\omega \hbar}  \ p_i$ and the position
$q_i \rightarrow \sqrt{ \hbar / \omega} \  q_i$
such that they have the same
units.  Then the volume of the full phase space is
given by

\be
Vol  = \int \prod_{j=1}^M \left( {dp_j dq^j \over \hbar} \right).
\label{vol1}
\ee

\noindent
We have the standard Cartesian metric,

\be
d \sigma ^2 = d {q_1}^2 + d {p_1}^2 + d {q_2}^2 + d {p_2}^2.
\ee

To find the metric of the reduced phase space, we will restrict
the coordinates to the constraint equation and choose the local
gauge fixing term ($\tan ^{-1} (p_2 / q_2) = {\rm constant}$).
This gauge  fixing is not global because it is ill defined at
$p_2 = q_2 =0$. In order to find the induced metric
on the reduced phase space,
it is easier to work in a set of two polar coordinates,

\bea
{r_1}^2 = {p_1}^2 + {q_1}^2  & \qquad  \qquad &
{r_2}^2 = {p_2}^2 + {q_2}^2
\nonumber \\[1ex]
\theta_1 = \tan ^{-1} \left({p_1 \over q_1} \right) & \qquad
\qquad &  \theta_2 = \tan ^{-1} \left({p_2 \over q_2} \right)
\eea

\noindent
The constraint equation \req{constr2}, after our rescaling and change
of coordinates, looks like

\be
S^2 = {r_1}^2 + {r_2}^2 = {2 E \over \omega \hbar}.
\label{const2}
\ee

\noindent
After apply the constraint ($r_1 dr_1 = - r_2 dr_2$)
and the gauge fixing ($d \theta_2 =0$), the metric
on the reduced phase space becomes

\be
d {\sigma'}^2 = \left( 1 - {{r_1}^2 \over S^2} \right) ^ {-1}
d {r_1}^2 + {r_1}^2 d {\theta_1}^2.
\ee

\noindent
We see that this metric is a constant curvature metric
($R = {2 / S^2}$).
The metric is ill defined at $r_1 = S$, which is also the same
place that the gauge fixing term is ill defined ($r_2 =
p_2 = q_2 =0 $).

We can
use this induced metric on the reduced phase space to
tell us a bit more about the system.
Let us follow a similar system that Klauder discussed
\cite{KlauderR}.\footnote{Note, the form of this metric is
slightly deferent then Klauder's. The results
is a difference of a factor of two in terms of the area.}
On the two sphere the total surface area must be
quantized in order that the term $\exp (i \oint p dq)$ for
a closed path be unambiguous. Note, we can not include the
possibility that $S=0$ because
the metric (and the gauge fixing) are ill
defined. Hence

\be
2 \pi n = \int dp \wedge dq =
\int \sqrt{g} \ dp dq = \pi S^2 \qquad n=1,2,3, \ldots
\label{quant1}
\ee

\noindent
This implies that the energy is also quantized
$E = \hbar \omega n$  where $n =1,2,3, \ldots$.
The reduced Hamiltonian is just zero so the resulting
propagator on the reduced phase space depends only on
the Weiner measure.  We can see this in the path integral

\be
{\cal N} \int D p_1 D q_1 e^{-{1 \over 2 \nu}
\int ({d \sigma ' \over dt})^2} e^{i \int p_1 dq_1}  .
\label{weiner2}
\ee

\noindent
Such a Weiner measure gives rise to spin-like
kinematical operators $S_i$ where $[S_i,S_j] = i \epsilon_{ijk}
S_k$ \cite{KlauderR}.
We will not work through the details of the resulting
spin system here because the details can be seen when
we consider the Dirac quantization.

Now we would like to consider the Dirac quantization of this
system (see \cite{Lawrie} for a similar discussion).
Let us begin by constructing the projection operator \req{proper}.
Similar to the single harmonic oscillator \req{conop1} the constraint
operator can be defined in terms of the raising and lowering
operators for the independent oscillators.

\be
\hat{\Phi} = a^\dagger a + b^\dagger b - E' ,
\qquad \qquad E' = E / \omega \hbar - 1.
\label{conop2}
\ee

\noindent
Because each oscillator is independent (before the constraint is
applied),
the double harmonics oscillator has a complete set of vectors that
is just the direct product of the eigenvalue of each of these
number operators,

\bea
&& |m,n \rangle = |m \rangle \otimes |n \rangle.
\\[1ex]
&& a^\dagger a | m,n \rangle = m | m,n \rangle,
\qquad
b^\dagger b | m,n \rangle = n | m, n \rangle .
\label{raise2}
\eea

\noindent
The constraint on this basis then quantizes the energy
$E = \hbar \omega (m + n + 1)$. In terms of the
Dirac quantization the energy would have to be an integer
($E = m' = 1,2,3, \dots$).  The resulting physical states
would be given by

\be
|\Psi \rangle_{phys} = |n, m'-n \rangle.
\label{phys5}
\ee

On this set of states, the raising and lowering operators
form above \req{raise2} are not defined, in that
$(a |\Phi \rangle _{phys})$ is not a physical state.
So, we need to find another set of operators that are
defined on this set of states.  These operators are equivalent
to the spin operators.

For convenience,
let us continue to use the scaled momentum and position
\req{vol1}.  In term of these coordinates,
we can define a new set of coordinates which have zero
Poisson brackets with the constraint $\{ s_i, \phi \} = 0$.
They are

\bea
s_1 & = & {1 \over 2} ( p_1 p_2 + q_1 q_2) ,
\label{coord4}
\\[1ex]
s_2 & = & {1 \over 2} ( p_2 q_1 - p_1 q_2) ,
\\[1ex]
s_3 & = & {1 \over 4} ( {p_1} ^2 + {q_1}^2 - {p_2}^2 - {q_2}^2) .
\eea

\noindent
This set of coordinates possess the standard $SO(3)$ Lie algebra,
$\{ s_i,s_j \} = \epsilon_{ijk} s_k$. The square of these three
coordinates is the constraint surface radius

\be
{s_1}^2 + {s_2}^2 + {s_3}^2 = s_0^2 ={1 \over 4} S^2 .
\label{square}
\ee

\noindent
Because the coordinates have a zero Poisson bracket with
the constraint, this set of coordinates is gauge invariant.
However, they are not all linearly independent, so let the reduced phase
space be described by $s_1,s_2$.
The induced metric is flat
${d \sigma'} ^2 = {s_0}^2({d s_1}^2 + {d s_2}^2)$, and the
domain is just a disk (${s_1}^2 + {s_2}^2 \le s_0^2$).

We can write a set of operators that correspond to the classical
coordinates above, eqs. \req{coord4} - \req{square}, that
preserves the $SO(3)$ algebra in terms
terms of our raising and lowering operators \req{raise2}.

\bea
\hat S_1 & = & {1 \over 2} ( a b^{\dagger} + a^{\dagger} b)
\qquad
\hat S_2  =  {i \over 2} ( a b^{\dagger} - a^{\dagger} b)
\nonumber \\[1ex]
\hat S_3 & = & {1 \over 2} ( a^{\dagger} a - b^{\dagger} b)
\qquad
\hat S_0  =  {1 \over 2} ( a^{\dagger} a + b^{\dagger} b)
\label{spinop}
\eea

\noindent
Then, it is possible to map the physical states give above
\req{phys5} onto the set of angular momentum eigenstates.

\bea
&& {{\hat S_0}}^2  | n, m'- n \rangle = {1 \over 4}
 m'(m'+1) | n, m'- n \rangle  \equiv j(j+1) |j,m \rangle ,
\\[1ex]
&& \hat S_3 | n, m' - n \rangle = {1 \over 2} (2n - m')
| n, m'- n \rangle  \equiv m |j,m \rangle .
\label{basis1}
\eea

\noindent
This means that $ j = 2m'$ and $ m = n- j$.  The raising and lowering
operators for the angular momentum
$\hat S_{\pm} = \hat S_1 \pm i \hat S_2$ act on this set of
states in the normal way,

\bea
&& \hat S_{+} |j,m \rangle = \sqrt{(j-m) (j+m+1) } | j , m+1 \rangle
\\[1ex]
&& \hat S_{-} |j,m \rangle = \sqrt{(j+m) (j-m+1) } | j, m-1 \rangle .
\label{rise1}
\eea

We can now construct the $SO(3)$ coherent states form these
operators \req{rise1} (see \cite{Perel2} for the details about this
coherent state). The coherent state is then

\be
| \xi \rangle = \exp \left( \xi \hat S_+
- \bar \xi \hat S_- \right) | \eta \rangle .
\label{cstate}
\ee

\noindent
The let us choose the lowest weight vector from above \req{rise1}
as our fiducial vector $ | \eta \rangle = |j, -j \rangle$.  Then
we can rewrite the above coherent state representation \req{cstate}
in terms of the above basis vectors \req{basis1}.

\be
| \xi \rangle = (1 + |\xi|^2) ^{-j} \sum_{m=-j}^j
\sqrt { 2j! \over (j+m)! (j-m)! } \ \xi ^{j+m} \ | j, m \rangle .
\label{sostate}
\ee

\noindent
This coherent state has a resolution of unity,

\be
\ID = \int {(2 j + 1) \over \pi}
{ d \xi d \bar \xi \over (1 + | \xi |^2)^2 }
| \xi \rangle \langle \xi |  .
\label{resol1}
\ee

\noindent
With this resolution of unity, we can construct the
path integral.  In the continuum limit, this path integral
appears as

\be
{\cal N } \int d {\mu'} _W ^{\nu}
\exp \left \{ { i \hbar \int { j \over (1 + |\xi|^2) }
\left( d \xi \bar \xi - \xi d \bar \xi \right)
} \right \} .
\ee

\noindent
The Weiner measure for this system was described in
\cite{Klauder5}.  We see that in fact the reduced
phase and the Dirac quantization result in a spin system,
where the energy of the system is mapped onto the total angular
momentum.

On first appearances would seem that the reduced phase space and
Dirac quantization lead to the same results. In both cases the
the energy is quantized $E= \omega \hbar (n +1)$ where
$n = 0,1,2,\ldots$ (see eqs. \ref{quant1}, \ref{phys5}).
However, this result is dependent
on the fact that we removed the point where the gauge
fixing went bad from the reduced phase space.
It is possible to include this point by using
two coordinate patches instead of one, where each coordinate patch has
its own gauge fixing.  Then, we can map this gauge fixing across the
boundary.  In so doing, we can include the zero energy term in the reduced
phase space \req{quant1}.  The result is that the two systems then have
different ground state energies \cite{Lawrie2}.

In the case of  constrained coherent state path integral, we will end up
integrating over the gauge orbits in effect averaging over
all possible gauge orbits.  Because of this, we will
not have to fix a gauge and we will not encounter this Gribov
problems.

Let us work through the projection operator
approach to the constrained coherent states for this system.
Extending the single state oscillator,
the coherent state for the double harmonic oscillator
can be written as

\be
|\alpha, \beta \rangle =
e^{-{|\alpha|^2 / 2} - {|\beta|^2 / 2} }
\sum _{m,n} ^{\infty} { 1 \over \sqrt{n!} \sqrt{m!}}
\alpha ^m \beta ^n  | m,n \rangle.
\ee

\noindent
Then we can project this on the physical states.

\bea
|\alpha, \beta \rangle_{phys} &= &
\int e^{i \lambda \hat{\Phi}} d \mu(\lambda) \left(
e^{-{|\alpha|^2 / 2} - {|\beta|^2 / 2} }
\sum _{m,n} ^{\infty} { 1 \over \sqrt{n!} \sqrt{m!}}
\alpha ^m \beta ^n  | m,n \rangle \right)
\nonumber \\[1ex]
& = & e^{-{|\alpha|^2 / 2} - {|\beta|^2 / 2} }
\sum _{m,n} ^{\infty} { 1 \over \sqrt{n!} \sqrt{m!}}
\alpha ^m \beta ^n \left(\int e^{i \lambda (n + m - E)}
d \mu(\lambda) \right) | m,n \rangle
\eea

\noindent
We will again choose Klauder's measure for non-compact
groups \req{uncmeas} for the measure for this projection.
Then similar to the single harmonic
oscillator \req{dfunc}, the physical vector is null unless
$E$ is arbitrarily close to an integer.  So let $E = m' = m+n$.
Then the physical vector is given by

\be
|\alpha, \beta \rangle_{phys} =
e^{-|\alpha|^2 - |\beta|^2 }
\sum _{n=0} ^{m'} \sqrt{ 1 \over n! (m'-n)!}
\alpha ^n \beta ^{m'-n}  | n, m'-n \rangle
\label{nphys1}
\ee

\noindent
Now we wish to normalize the physical vector.  It is just a
quick calculation to show that the normalized physical vector
is

\be
| \alpha, \beta \rangle_{phys} = \left( | \alpha|^2 +
|\beta|^2 \right) ^{-{m' \over 2}}
\sum _{n=0} ^{m'} \sqrt{ m'! \over n! (m'-n)!}
\alpha ^n \beta ^{m'-n}  | n, m'-n \rangle
\ee

It is easy to see that the gauge transformation generated by the
constraint is $\alpha \rightarrow \alpha e^{i\theta}$ and
$\beta \rightarrow \beta  e^{i\theta}$.  Like the single
harmonic oscillator \req{single} this gauge transformation appears as
an overall phase in front of the physical vector

\be
|\alpha , \beta  \rangle _{phys} \rightarrow e^{im' \theta}
|\alpha , \beta \rangle _{phys}
\ee

\noindent
To remove the gauge dependence, let us then
define $\xi = \alpha / \beta$.  Writing the physical state
in terms of this variable, we can factor out the gauge
transformations which appear again as phase factor \req{single}.

\bea
| \alpha, \beta \rangle_{phys} & = &
\left( {\beta \over |\beta|} \right)^{m'}
\left( 1 + \left| {\alpha \over \beta} \right| ^2
 \right)^{- {m' \over 2}}
\sum _{n=0} ^{m'} \sqrt{ m'! \over n! (m'-n)!}
\left( {\alpha \over \beta} \right) ^ n  | n, m'-n \rangle
\nonumber \\[1ex]
& = & e^{ i m' \theta} \left( 1 + |\xi|^2 \right)^{- {m' \over 2}}
\sum _{n=0} ^{m'} \sqrt{ m'! \over n! (m'-n)!}
\ \xi ^ n \ | n, m'-n \rangle .
\label{phase1}
\eea

\noindent
It is easy to see that the physical coherent state maps onto
the $SO(3)$ coherent state \req{sostate}.  The energy is mapped onto
total angular momentum $j= 2 E'$.

\be
| \xi \rangle = (1 + |\xi|^2) ^{-j} \sum_{m=-j}^j
\sqrt { 2j! \over (j+m)! (j-m)! } \ \xi ^{j+m} \ | j, m \rangle
\label{phys1}
\ee

\noindent
The resulting reduced phase space agrees with the information that
we were able to discern from the reduced phase space coherent state
discussed earlier \req{weiner2}.  It is clearly a spin system with
the total angular momentum given by the the energy ($ 2j = E'$).

The propagator for this system is simply the overlap function
of the $SO(3)$ coherent state,

\bea
\langle \xi' | \xi \rangle & \equiv &
\langle \alpha', \beta' | \IP | \alpha \beta \rangle
\nonumber \\[1ex]
& = & ( 1 + |\xi'|^2) ^{-j} ( 1 + |\xi|^2 )^{-j}
( 1 + \bar \xi ' \xi)^{2j} .
\label{dyn1}
\eea

\noindent
At any given ``time'', the coherent state gives
a minimum uncertainty wave packet $\langle {J_1}^2 \rangle
\langle {J_2}^2 \rangle = {1 \over 4} \langle J_0 \rangle ^2 $.
The most probable matrix element (the classical solution) is
simply $\xi' = \xi$.  Any expectation value
of the spin operators \req{spinop} is simple enough to
calculate as well. From these expectation value
the classical ``dynamics'' of the reduced phase space can be
can deduced.

Returning to the full phase space, we would like to reconstruct the
the classical equations of motion as we did in the single oscillator
case \req{eqmotion4}.  Using the non-normalized physical vector
\req{nphys1}, we can define the physical state on the full
phase space as

\be
| \phi_{ \alpha' \beta'} \rangle =
 \IP | \alpha ' , \beta ' \rangle
\ee

The wave function of this state on the phase space is again the
overlap function,

\bea
\phi_{\alpha' \beta' } (\alpha '' , \beta '' ) & = &
\langle \alpha '', \beta '' | \phi_{ \alpha ' , \beta '} \rangle
\nonumber \\[1ex]
& = & e ^ { -{1 \over 2} ( |\alpha ''| ^2  + |\beta''| ^2
+ |\alpha '| ^2 + |\beta '| ^2 )} \
{( {\bar \alpha}'' \alpha ' + {\bar \beta }''
\beta ' ) ^m \over m!}
\eea

\noindent
We can renormalize the state such that at its peek, it is
approximately one, as we did in the single oscillator case.
Let $\alpha ' = r' e^{i \theta'}, \beta' = \rho' e^{i \phi'},$ etc.
Then,

\bea
&& \Big| \langle \alpha '', \beta '' | \phi_{ \alpha'  \beta '}
\rangle \Big| = \nonumber \\[1ex]
&& \sqrt{2 \pi m} \ e ^ { -{1 \over 2} ( | r''| ^2  + |\rho''| ^2
+ | r '| ^2 + | \rho '| ^2 )} {
\Big( {r'} ^2 {r'' }^2 + {\rho ' }^2 {\rho ''}^2 +
2 r' r'' \rho ' \rho '' \cos(\theta) \Big) ^{m}
\over m!}
\eea

\noindent
where $\theta = \theta' - \theta'' - \phi' + \phi ''$. The peak of
this function is at $ \theta = 0, r' = r'', \rho ' = \rho '',$ and
${r'}^2 + {\rho '} ^2 = m$.
Note that $\theta = 0$ implies that the states evolve together
as they do in the classical equation of motion \req{eqmotion3}.
The correlation functions
of the the position and momentum are give by

\bea
\langle \alpha '' , \beta ''| \hat Q_1 | \phi_{\alpha ' \beta '} \rangle
& = &  \sqrt{ 2 \hbar \over \omega} \left(  {
m \alpha' \over ({\bar \alpha}'' \alpha ' + {\bar \beta}'' \beta ')}
+ \bar \alpha '' \right)
\langle \alpha '' , \beta ''| \phi_{\alpha ' \beta '} \rangle,
\\[1ex]
 \langle \alpha '' , \beta ''| \hat P_1
| \phi_{\alpha ' \beta '} \rangle
& = &  i \sqrt{ 2 \hbar \omega} \left( {
m \alpha' \over ({\bar \alpha}'' \alpha ' + {\bar \beta}'' \beta ')}
- \bar \alpha '' \right)
\langle \alpha '' , \beta ''| \phi_{\alpha ' \beta '} \rangle .
\label{corr2}
\eea

\noindent
Likewise for the position and momentum of the second oscillator.
Expanding about the peak, we have

\bea
&&\langle \alpha'', \beta'' |\hat Q_1
|\phi_{\alpha ' \beta '} \rangle = { r' \over \omega}
\cos(\theta' - \theta'') + {\cal O}(\hbar)
\\[1ex]
&&\langle \alpha'', \beta''| \hat P_1 | \phi_{\alpha ' \beta '}
\rangle = { r'} \sin(\theta' - \theta'') + {\cal O}(\hbar)
\\[1ex]
&& \langle \alpha '' , \beta ''| \hat Q_2
| \phi_{\alpha ' \beta '} \rangle = { \rho ' \over \omega}
\cos(\phi' - \phi'') + {\cal O}(\hbar)
\\[1ex]
&& \langle \alpha '' , \beta '' | \hat P_1
|\phi_{\alpha ' \beta '} \rangle = { \rho'}
\sin(\phi' - \phi'') + {\cal O}(\hbar)
\eea

\noindent
The  energy for each of the oscillators in the classical limit is

\bea
&& \langle \phi_{\alpha ' \beta '} |
\hat E_1 | \phi_{\alpha ' \beta '} \rangle = {r'}^2 + {\cal O}(\hbar),
\qquad \qquad
\langle \phi_{\alpha ' \beta '} | \hat E_2
| \phi_{\alpha ' \beta '} \rangle = {\rho '}^2 + {\cal O}(\hbar),
\nonumber \\[1ex]
&& \langle \phi_{\alpha ' \beta '} | E_{total}
| \phi_{\alpha ' \beta '} \rangle = m'
= {r'}^2 + {\rho '}^2 + {\cal O}(\hbar)
\eea

\noindent
So, we see that we meet the constraint equation \req{constr3},
and we get back the classical
equations of motion \req{eqmotion3} in the classical limit.

\section{Discussion}

It is interesting to note that the gauge degree of freedom for both
of these system comes out in a phase factor in front of the
reduced phase space coherent state. Although is might just be a
artifact of the harmonic oscillator(s), it is suggestive that the
gauge should appear in this way in general.  If this is the case,
the only dependence on the gauge orbits in the path integral
will appear in the one form,

\bea
\langle p,q| {d \over dt} | p,q \rangle dt & = & p dq
\nonumber \\[1ex]
\langle p,q | e^{-i f(p,q)} {d \over dt}  e^{i f(p,q)}
| p,q \rangle dt & = & p dq + df
\eea

\noindent
If there is not a boundary, this difference can just be integrated out
as a total derivative.  If there is a boundary, we pick up a boundary
term that is still dependent on the gauge orbits.  For example,
such a gauge symmetry
break term is seen in the relationship between the Chern-Simons actions
and the Wess-Zumino-Witten action.

It should be noted that, the above correlations functions
\req{corr1} and \req{corr2} are not physical observables.  This correlations
are still dependent on the gauge degree of freedom.  However,
they do show that the equations of motions are still embedded
in the formulation of the coherent state on the reduced phase space.
In terms of the double harmonics oscillator, it is interesting
to note that the
width of the correlation function in the angular direction is
dependent on the energy of each oscillators,

\be
\sigma \sim { E_{total} \over E_1 E_2}.
\ee

\noindent
In terms of ``quantum clocks,'' this means that at low energies, the
correlations between this clocks may become fuzzy.  Certainly, a
more precise statement in terms of observables needs to be considered.

\section{Acknowledgments}

I would like to thank Steve Carlip for all of his support and time.
I would also like to thank Richard Epp for useful discussions,
John Klauder for his comments.
This work was supported by National Science Foundation grant PHY-93-57203
and Department of Energy grant DE-FG03-91ER40674.

\end{document}